# Topological structures of energy flow: Poynting vector skyrmions

Sicong Wang[1], Jialin Sun[1], Zecan Zheng[1], Zhikai Zhou[1], Hongkun Cao[1], Shichao Song[1], Zi-Lan Deng[1], Fei Qin[1], Yaoyu Cao[1], and Xiangping Li[1*]

[1] *Guangdong Provincial Key Laboratory of Optical Fiber Sensing and Communications, Institute of Photonics Technology, Jinan University, Guangzhou 510632, China*

Topological properties of energy flow of light are fundamentally interesting and have rich practical applications in optical manipulations. Here, skyrmion-like structures formed by Poynting vectors are unveiled in the focal region of a pair of counter-propagating cylindrical vector vortex beams in free space. A Néel-Bloch-Néel skyrmion type transformation of Poynting vectors is observed along the light propagating direction within a volume with subwavelength feature sizes. The corresponding skyrmion type can be determined by the phase singularities of the individual components of the coherently superposed electromagnetic field in the focal region. This work reveals a new family member of optical skyrmions and may introduce novel physical phenomena associated with light scattering and optical force.

Topological photonics inspired by the discovery of photonic topological insulators [1-5] opens a path towards the discovery of fundamentally new states of light [6,7] and potential revolutionary applications, such as topologically protected lasing [8-10], photonic circuitry and slow light [11]. Skyrmions, originally proposed by Skyrme in 1962 [12] and acting as a typical branch of topologically nontrivial textures, were once demonstrated in Bose-Einstein condensates [13], nematic liquid crystals [14], and as a phase transition in chiral magnets [15,16]. Especially, magnetic skyrmion lattice and single magnetic skyrmions have been extensively studied and are considered as a promising route towards high-density magnetic information storage, transfer, and spintronic devices [17-23]. Recently, optical skyrmions or plasmonic skyrmions emerge as fire-new objects of study in topological photonics and expand the family of skyrmions [24-27]. Néel-type optical skyrmion lattices formed by the electric field vectors are generated by the interference of the propagating surface plasmons excited through the hexagonal grating structures on a metal surface [24,25]. Meanwhile and likewise, resorting to the interference of the evanescent surface plasmons, a single Néel-type optical skyrmion formed by spin angular momentum vectors is generated by tightly focusing a cylindrical vector vortex beam on a metal surface and variations of the local spin direction on a deep-subwavelength scale down to 1/60 $\lambda$ are demonstrated [26]. Subsequently, more optical skyrmions structured by field or spin vectors are proposed and enrich this new field of topological photonics promptly [27-34]. These pioneering works may facilitate stunning applications in optical storage, high-resolution imaging, and precision metrology. Being a new cutting-edge study in topological photonics, trend of skyrmion research is better to be further rendered.

As another significant natural property of light, energy flow, as the name suggests, describes the flow behavior of the energy of light. It plays an important role in understanding the physics of light scattering [35] and optical force [36,37]. With the development of vector beams and tight focusing manner, complex behaviors of Poynting vectors are unveiled under focal conditions. Transversal toroidal energy flow [38-41], longitudinal energy backflow [42-47], and some fundamental topological structures of Poynting vectors [48] have been studied in detail. However, whether energy flow can behave as a skyrmion-like manner remains elusive.

In this paper, optical skyrmions formed by Poynting vectors are unveiled in the focal region of a pair of counter-propagating cylindrical vector vortex beams in free space. Theoretical analysis manifests that the Poynting vectors experience a Néel-Bloch-Néel skyrmion type transformation along the light propagating direction within a volume with subwavelength feature sizes, and the corresponding skyrmion type is tightly correlated with the phase singularities of the individual components of the coherently superposed electromagnetic field in the focal region.

As shown in Fig. 1, a pair of counter-propagating cylindrical vector vortex beams are focused through a $4\pi$ microscopic configuration. The polarization and the phase distribution of each incident cylindrical vector vortex beam can be designed and analyzed by dividing itself into two parts, one with azimuthally polarized electric field **E** and radially polarized magnetic field **H** (Part I or Part III), and the other one with radially polarized electric field **E**, azimuthally polarized magnetic field **H** and a phase delay of $\pi/2$ supplied by a homogeneous phase plate (PP1/PP2) (Part II or Part IV). For simplicity, in our work, the four parts are assumed to have equivalent homogeneously distributed light intensity, except for the polarization singularity in the center where the light intensity is zero. **E**, **H** and their corresponding propagating direction, namely the wave vector, satisfy the right-handed spiral relationship. Vortex phase plates (VPP1 and VPP2) with topological charges of

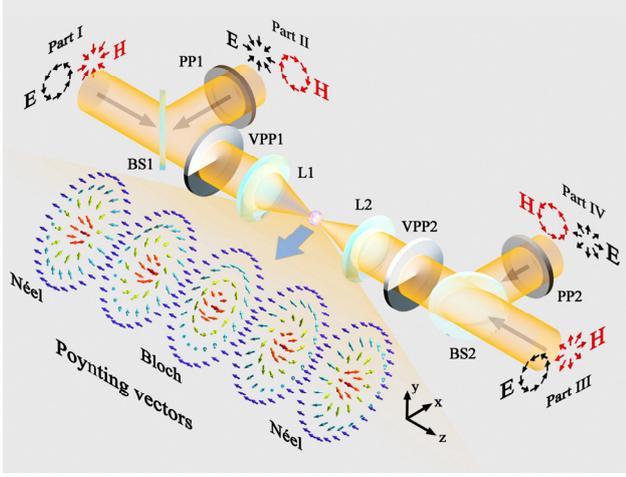

FIG. 1. Schematic of the skyrmion type transformation of the Poynting-vector structures in the focal region of a pair of counter-propagating cylindrical vector vortex beams. PP: homogeneous phase plate with a phase delay of $\pi/2$, BS: beam splitter, VPP: first-order vortex phase plate, L: focal lens, **E**: incident electric field, **H**: the corresponding incident magnetic field.

+1 or -1 are used to introduce first-order vortex phases. The orientations of the **E** and the **H** vectors indicate that, under this $4\pi$ focal condition, the transversal and the longitudinal components of the coherently superposed electromagnetic field in the focal plane can be modulated independently by the azimuthally polarized and the radially polarized **E** or **H** vectors with vortex phases, respectively. The enlarged view of the focal spot denotes the schematic of the skyrmion type transformation of the Poynting-vector structures along the light propagating direction.

According to the Richards and Wolf diffraction theory [49] and the characteristics of the field superposition in the focal region of a $4\pi$ microscopic configuration [50-52], the electromagnetic field in the focal region can be expressed as

$$\mathbf{E}_f(r,\varphi,z) = A e^{i\varphi} \int_0^{\theta_{\max}} \begin{bmatrix} V_1 W_1 + V_2 W_2 \cos\theta \\ i(V_2 W_1 + V_1 W_2 \cos\theta) \\ -2i V_3 W_1 \sin\theta \end{bmatrix} \begin{matrix} \mathbf{e}_r \\ \mathbf{e}_\varphi \\ \mathbf{e}_z \end{matrix} \cdot T d\theta, \quad (1)$$

$$\mathbf{H}_f(r,\varphi,z) = -\frac{iA}{\mu_0 c} e^{i\varphi} \int_0^{\theta_{\max}} \begin{bmatrix} V_1 W_1 + V_2 W_2 \cos\theta \\ i(V_2 W_1 + V_1 W_2 \cos\theta) \\ -2i V_3 W_1 \sin\theta \end{bmatrix} \begin{matrix} \mathbf{e}_r \\ \mathbf{e}_\varphi \\ \mathbf{e}_z \end{matrix} \cdot T d\theta, \quad (2)$$

where

$$\begin{cases} V_1 = J_0(kr\sin\theta) + J_2(kr\sin\theta) \\ V_2 = J_0(kr\sin\theta) - J_2(kr\sin\theta), \\ V_3 = J_1(kr\sin\theta) \end{cases} \quad (3)$$

and

$$\begin{cases} W_1 = e^{ikz\cos\theta} + e^{-ikz\cos\theta} \\ W_2 = e^{ikz\cos\theta} - e^{-ikz\cos\theta} \\ T = \sqrt{\cos\theta}\sin\theta \end{cases} \quad (4)$$

$r$, $\varphi$, and $z$ are the cylindrical coordinates in the focal region. $\theta$ is the converging angle, which varies from 0 to $\theta_{\max}$. $\theta_{\max}$ = arcsin(NA) is the maximum converging angle of the focal lenses and NA = 0.95 is the numerical aperture of the focal lenses. $A$ is a complex constant. $\mu_0$ is the permeability of vacuum and $c$ is the speed of light in vacuum. $\mathbf{e}_r$, $\mathbf{e}_\varphi$, and $\mathbf{e}_z$ are the unit base vectors. $k$ is the wave vector of the beams. $J_0$, $J_1$, and $J_2$ denote the Bessel functions of the first kind. More detailed formula derivations are shown in Supplementary Note 1.

From Eqs. (1) and (2), it can be seen that, except for a phase delay of $3\pi/2$ or a phase lead of $\pi/2$, the focal magnetic field $\mathbf{H}_f$ has the same normalized field distributions with the focal electric field $\mathbf{E}_f$. Figure 2 shows the normalized amplitude and the phase distributions of the focal electromagnetic field within the $x$-$z$ plane including the ray axis. In this plane, $\mathbf{E}_r$ ($\mathbf{H}_r$) and $\mathbf{E}_\varphi$ ($\mathbf{H}_\varphi$) are substituted by $\mathbf{E}_x$ ($\mathbf{H}_x$) and $\mathbf{E}_y$ ($\mathbf{H}_y$) to provide better field presentation. It is worth noting that phase singularities emerge for all of the field components and result in zero amplitude distributions in the corresponding regions. Phase singularities are ubiquitous phenomena in the interference of multiple electromagnetic waves and may give an additional insight into propagation and interaction of electromagnetic waves [36]. In our case, through the interference of a pair of counter-propagating cylindrical vector vortex beams, different singularity distributions are formed for different field components. For the $x$ component, as shown in Fig. 2(b), four main phase singularities with topological charges of $m = \pm 1$ at $z = \pm 0.35\lambda$, denoted by the red circles, can be seen. $\lambda$ is the wavelength of the light beams. At these singular points, $\mathbf{E}_x$ ($\mathbf{E}_r$) and $\mathbf{H}_x$ ($\mathbf{H}_r$) vanish and the other components may result in Poynting vectors with radial orientations. Similar vortex phase singularities can be seen at $z = 0$ for the $y$ component as shown in Fig. 2(d). At these points, $\mathbf{E}_y$ ($\mathbf{E}_\varphi$) and $\mathbf{H}_y$ ($\mathbf{H}_\varphi$) vanish and the other components may result in Poynting vectors with azimuthal orientations. Regarding the $z$ component, binary phase ($\pi/2$ and -$\pi/2$) distributions can be seen as shown in Fig. 2(f). Along the singular lines between the yellow and the blue regions, $\mathbf{E}_z$ and $\mathbf{H}_z$ vanish and longitudinally oriented Poynting vectors can be predicted. The normalized amplitude and the phase distributions of the individual field components in the focal plane are shown in Supplementary Fig. S2. Exactly as indicated by Eqs. (1) and (2), all of the field components are encoded with a vortex phase with a topological charge of 1. In the main-lobe regions, $\mathbf{E}_\varphi$ ($\mathbf{H}_\varphi$) and $\mathbf{E}_z$ ($\mathbf{H}_z$) have a phase delay and a phase lead of $\pi/2$ compared with $\mathbf{E}_r$ ($\mathbf{H}_r$), respectively. The polarization in the center is left-handed circular polarization, which has been discussed in Ref. [53].The phase distributions of the individual field components in the *x-y* plane at $z = \pm 0.21\lambda$ and $z = \pm 0.35\lambda$ are shown in Supplementary Fig. S3. It can be seen that the vortex phases of $\mathbf{E}_r$ ($\mathbf{H}_r$) and $\mathbf{E}_\varphi$ ($\mathbf{H}_\varphi$) rotate clockwise from $z = -0.35\lambda$ to $z = 0.35\lambda$ while those of $\mathbf{E}_z$ and $\mathbf{H}_z$ remain unchanged. From $z = 0$ to $z = \pm 0.35\lambda$, the vortex phases of $\mathbf{E}_r$ ($\mathbf{H}_r$) and $\mathbf{E}_\varphi$ ($\mathbf{H}_\varphi$) experience a change of $\pi/2$. This changing regular may boost the skyrmion type transformation of the Poynting vectors along the light propagating direction. The red circles depicted in Supplementary Fig. S2 and S3 indicate the regions within which Poynting vector skyrmions will be formed. The energy density of the electromagnetic field in the focal region are shown in Supplementary Fig. S4.

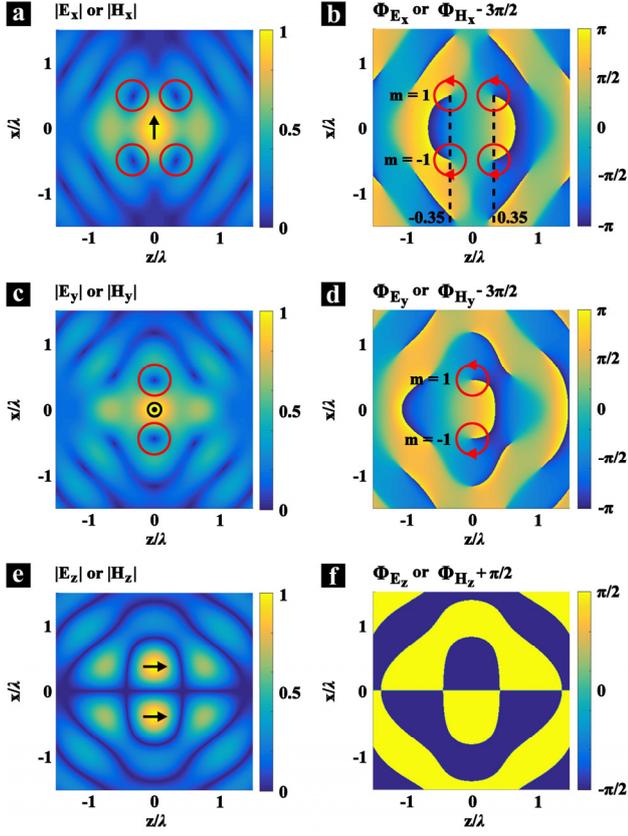

FIG. 2. The normalized amplitude and the phase distributions of the focal electromagnetic field within the *x-z* plane including the ray axis. (a), (c), and (e) The normalized amplitude distributions of $\mathbf{E}_x$ ($\mathbf{H}_x$), $\mathbf{E}_y$ ($\mathbf{H}_y$), and $\mathbf{E}_z$ ($\mathbf{H}_z$). The black arrows and dots denote the corresponding polarization orientations. (b), (d), and (f) The phase distributions of $\mathbf{E}_x$ ($\mathbf{H}_x$), $\mathbf{E}_y$ ($\mathbf{H}_y$), and $\mathbf{E}_z$ ($\mathbf{H}_z$). The red circles indicate the vortex phase singularities.

Subsequently, the Poynting vector, namely the energy flux density, of the electromagnetic field in the focal region can be calculated through

$$\mathbf{P}(r,\varphi,z) = \frac{1}{2}\mathrm{Re}(\mathbf{E}_f \times \mathbf{H}_f^*), \quad (5)$$

where $\mathbf{H}_f^*$ is the conjugate of $\mathbf{H}_f$. Especially, in the focal plane, Eq. (5) can be expressed as

$$\mathbf{P}(r,\varphi,0) = \frac{4|A|^2}{\mu_0 c}\begin{bmatrix} 0 \\ 2\int_0^{\theta_{max}} V_3 \sin\theta T d\theta \cdot \int_0^{\theta_{max}} V_1 T d\theta \\ \int_0^{\theta_{max}} V_1 T d\theta \cdot \int_0^{\theta_{max}} V_2 T d\theta \end{bmatrix}\begin{matrix}\mathbf{e}_r\\ \mathbf{e}_\varphi\\ \mathbf{e}_z\end{matrix}. \quad (6)$$

Eq. (6) indicates that only the azimuthal and the longitudinal components of the Poynting vectors exist in the focal plane. The corresponding normalized cross sections of these components are illustrated in Fig. 3(c). The whole picture of the generated Bloch-type Poynting vector skyrmion is shown in Fig. 3(a) and 3(b). Figure 3(a) shows the normalized amplitude and the unit vector projections of the Poynting vectors on the focal plane. The black arrows present an anticlockwise flow circulation. Within the red circle, a complete skyrmion structure is formed. Its three-dimensional vectorial structure formed by unit vectors is shown in Fig. 3(b). It can be seen that the Poynting vector reverses from the "up" state in the center to the "down" state at the periphery with a Bloch-type manner. $\theta_{xy}$ is the orientation angle of the Poynting vector with respect to the *x-y* plane. More quantificationally, Fig. 3(e) shows the variation of $\theta_{xy}$ along the radial direction from the center to the periphery. It illustrates that $\theta_{xy}$ varies monotonically from $\pi/2$ at $r = 0$ to $-\pi/2$ at $r = 0.824\lambda$ shown as the pink shadow area. Beyond this area, $\theta_{xy}$ still varies between $\pi/2$ and $-\pi/2$ at least within $r = 2\lambda$ where the amplitude of the Poynting vector is far less than that in the center. The corresponding three-dimensional vectorial structure formed by unit vectors along the radial direction is shown in Fig. 3(d). It can be seen that the Poynting vector varies its orientation from the "up" state to the "down" state continuously and repeatedly.

Skyrmion number ($N_{sk}$) and skyrmion number density are used to characterizing Skyrmion structures. The Skyrmion number can be written in an integral form [20,25,26]

$$N_{sk} = \frac{1}{4\pi}\iint_S n dS = \frac{1}{4\pi}\iint \mathbf{e}_p \cdot \left(\frac{\partial \mathbf{e}_p}{\partial x} \times \frac{\partial \mathbf{e}_p}{\partial y}\right) dxdy, \quad (7)$$

where the area *S* covers the complete Poynting vector skyrmion and *n* is the skyrmion number density. $\mathbf{e}_p$ represents the unit vector in the direction of the local Poynting vector. Figure 3(f) shows the skyrmion number density distribution of the Poynting vector skyrmion in the focal plane and the corresponding calculated skyrmion

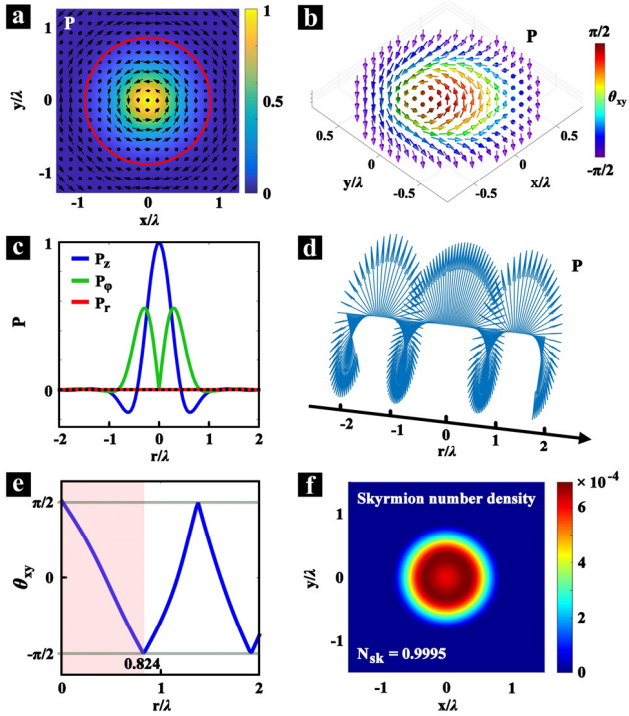
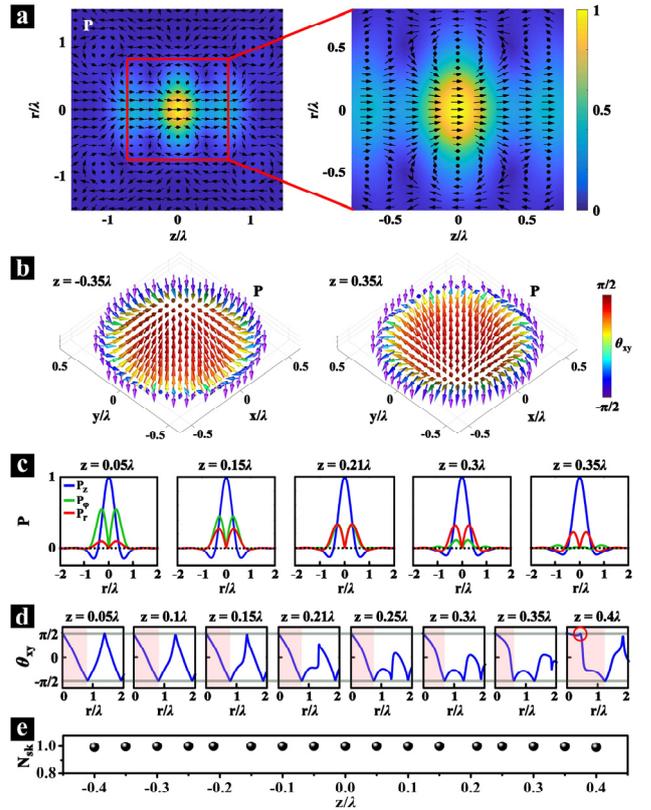

FIG. 3. Distributions of the Bloch-type Poynting vector skyrmion in the focal plane. (a) The normalized amplitude and the unit vector projections of the Poynting vectors on the focal plane. (b) The three-dimensional vectorial structure of the Poynting vector skyrmion formed by unit vectors within the red circle in (a). (c) The normalized cross sections of the individual components of the Poynting vector skyrmion. (d) The three-dimensional vectorial structure formed by unit vectors along the radial direction through the center. (e) Variations of $\theta_{xy}$ versus $r$. (f) The skyrmion number and the skyrmion number density distribution of the Poynting vector skyrmion in the focal plane.

number equals 0.9995, the deviation of which from 1 stems from the calculation error.

The normalized amplitude and the unit vector projections of the Poynting vectors on the $r$-$z$ plane are shown in Fig.4(a). From the enlarged area indicated by the red square, changes of the azimuthal component of the Poynting vectors, namely the transformation between the black arrows and the black dots along the $z$ direction, can be seen. This changing regular may imply a type transformation of the Poynting vector skyrmion along the light propagating direction. Figure 4(b) shows the three-dimensional vectorial Poynting-vector structures formed by unit vectors at $z = \pm 0.35\lambda$. At both positions, the Poynting vector reverses from the "up" state in the center to the "down" state at the periphery with a Néel-type manner. The only difference is that the Néel-type structures present inward-pointing and outward-pointing configurations at $z = -0.35\lambda$ and $z = 0.35\lambda$, respectively. The corresponding normalized amplitude and the unit vector projections on the $x$-$y$ plane and on the $r$-$z$ plane are shown in Supplementary Figs. S5(a) and S5(b), respectively.

FIG. 4. Skyrmion type transformations of the Poynting vectors along the light propagating direction. (a) The normalized amplitude and the unit vector projections of the Poynting vectors on the $r$-$z$ plane. (b) The three-dimensional vectorial structures of the Poynting vectors formed by unit vectors at $z = \pm 0.35\lambda$. (c) The normalized cross sections of the individual components of the Poynting vectors at different longitudinal positions. (d) Variations of $\theta_{xy}$ versus $r$ at different longitudinal positions. (e) Skyrmion numbers at different longitudinal positions.

More quantificationally, Fig. 4(c) combined with Supplementary Fig. S6(a) illustrates the normalized cross sections of the individual components of the Poynting vectors at different longitudinal positions. It can be seen that the azimuthal component $P_\varphi$ decreases gradually and the radial component $P_r$ emerges from $z = 0$ to $z = \pm 0.35\lambda$. At $z = \pm 0.35\lambda$, $P_\varphi$ is almost vanishing in the central region. The corresponding purity of the local Poynting vector components in the $r$-$z$ plane ($P_r$ and $P_z$), calculated by $(|P_r|^2+|P_z|^2)/(|P_r|^2+|P_\varphi|^2+|P_z|^2)$ and shown in Supplementary Fig. S4(c), remains higher than 97% within the whole skyrmion region.

Figure 4(d) combined with Supplementary Fig. S6(b) illustrates the variations of $\theta_{xy}$ along the radial direction from the center to the periphery at different longitudinal positions. $\theta_{xy}$ varies between $\pi/2$ and $-\pi/2$ continuously and repeatedly within $r = 2\lambda$ at longitudinal positions where $|z| \leq 0.15\lambda$. When $0.21\lambda \leq |z| \leq 0.35\lambda$, $\theta_{xy}$ varies monotonically from $\pi/2$ to $-\pi/2$ only within a certain range indicated by the pink shadow. If $|z| > 0.35\lambda$, for example $|z| = 0.4\lambda$, $\theta_{xy}$ does not vary monotonically any more as indicated by the red circles and hence breaks the skyrmion structure. When $0 < |z| < 0.35\lambda$, Poynting vector skyrmions with a hybrid type, namely the superposition of the Neel type and the Bloch type with non-zero $P_r$, $P_\varphi$, and $P_z$ components, can be formed. Figures 5(a)-5(d) show the hybrid-type skyrmion structures of the Poynting vectors at $z = \pm 0.21\lambda$ where $|P_r| = |P_\varphi|$ as indicated in Fig. 4(c) and Supplementary Fig. S6(a).

The skyrmion numbers within the range $|z| \leq 0.35\lambda$ are calculated and shown in Fig. 4(e), all of which are almost equal to 1. Although the skyrmion number does not change within this longitudinal range, the skyrmion number density varies significantly. The skyrmion number densities at $z = \pm 0.21\lambda$ and $z = \pm 0.35\lambda$ are shown in Figs. 5(d) and 5(e), respectively. Compared with Fig. 3(f), it can be seen that from $z = 0$ to $z = \pm 0.35\lambda$ the skyrmion number density is gradually "pushed" outwards from the center, which signifies that at longitudinal positions far from the focal plane, the Poynting vectors change their orientations from the "up" state to the "down" state intensely and sharply in the peripheral regions of the skyrmion structures.

It is worth noting that the spin vectors of the focal electromagnetic field under this $4\pi$ focal condition also exhibit similar skyrmion structures and transformation. This can be well understood by analyzing Eqs. (1) and (2). As except for a phase delay of $3\pi/2$ or a phase lead of $\pi/2$, $\mathbf{H}_f$ has the same normalized field distributions with $\mathbf{E}_f$, the spin density of the electric field $\mathbf{S}_E \propto \mathrm{Im}(\mathbf{E}^* \times \mathbf{E})$, the spin density of the magnetic field $\mathbf{S}_H \propto \mathrm{Im}(\mathbf{H}^* \times \mathbf{H})$, and the Poynting vector should also have identical normalized distributions.

In summary, as a new family member of the optical skyrmions, Poynting vector skyrmions are constructed in the focal region of a pair of counter-propagating cylindrical vector vortex beams in free space. Through analyzing the variations of the vortex phases and the phase singularities of the individual components of the focal electromagnetic field along the light propagating direction, a Néel-Bloch-Néel skyrmion type transformation of the Poynting vectors within a volume with subwavelength feature sizes have been observed. The handedness of the Bloch-type Poynting vector skyrmion can be reversed by introducing an additional phase of $\pi$ to Part II and Part IV. We envision that these theoretical results may expand the research scope of topological photonics and provide guidance to realize novel optical manipulations associated with light scattering and optical force.

This research is financially supported by the National Key R&D Program of China (2021YFB2802003), National Natural Science Foundation of China (NSFC) (61975066, 62075084, 62075085, 62005104), Guangdong Basic and Applied Basic Research Foundation (2021A1515011586, 2020A1515010615, 2020B1515020058), Guangzhou Science and Technology Program (202002030258).

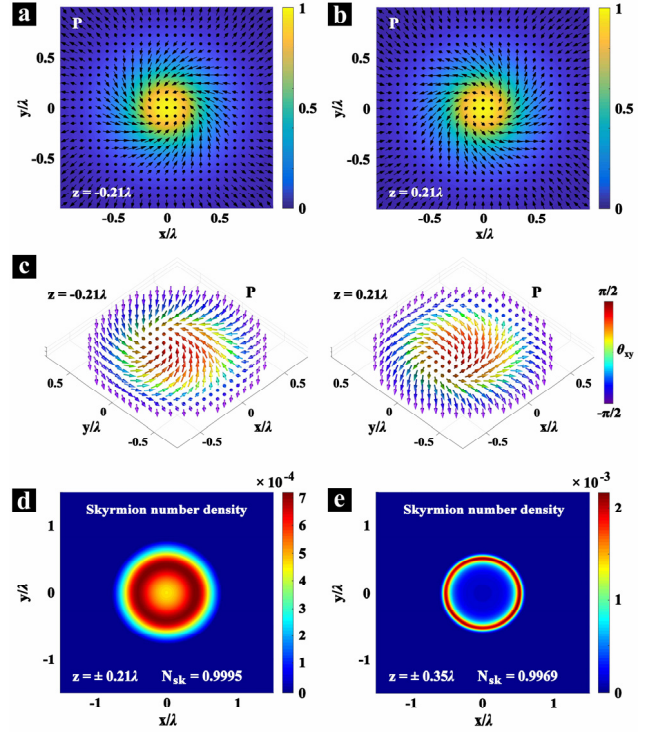

FIG. 5. (a) and (b) The normalized amplitude and the unit vector projections of the Poynting vectors on the $x$-$y$ plane at $z = \pm 0.21\lambda$. (c) The three-dimensional vectorial structures of the Poynting vectors formed by unit vectors at $z = \pm 0.21\lambda$. (d) and (e) The skyrmion numbers and the skyrmion number density distributions of the Poynting vector skyrmions in the $x$-$y$ plane at $z = \pm 0.21\lambda$ and $z = \pm 0.35\lambda$.

*Corresponding author.
xiangpingli@jnu.edu.cn